# Referenced Publication Year Spectroscopy (RPYS) and Algorithmic Historiography:

## The Bibliometric Reconstruction of András Schubert's Œuvre

Loet Leydesdorff,*[a] Lutz Bornmann,[b] Jordan Comins,[c] Werner Marx,[d] and Andreas Thor[e]

[a] * corresponding author; University of Amsterdam, Amsterdam School of Communication Research (ASCoR), PO Box 15793, 1001 NG Amsterdam, The Netherlands; email: loet@leydesdorff.net
[b] Division for Science and Innovation Studies, Administrative Headquarters of the Max Planck Society, Hofgartenstr. 8, 80539 Munich, Germany; email: bornmann@gv.mpg.de
[c] Center for Applied Information Science, Virginia Tech Applied Research Corporation, Arlington, VA, United States; email: jcomins@gmail.com
[d] Max Planck Institute for Solid State Research, Information Service, Heisenbergstrasse 1, 70506 Stuttgart, Germany; email: w.marx@fkf.mpg.de
[e] University of Applied Sciences for Telecommunications Leipzig, Gustav-Freytag-Str. 43-45, 04277 Leipzig, Germany; email: thor@hft-leipzig.de

**Abstract**

Referenced Publication Year Spectroscopy (RPYS) was recently introduced as a method to analyze the historical roots of research fields and groups or institutions. RPYS maps the distribution of the publication years of the cited references in a document set. In this study, we apply this methodology to the œuvre of an individual researcher on the occasion of a *Festschrift* for András Schubert's 70[th] birthday. We discuss the different options of RPYS in relation to one another (e.g. Multi-RPYS), and in relation to the longer-term research program of algorithmic historiography (e.g., *HistCite™*) based on Schubert's publications (n=172) and cited references therein as a bibliographic domain in scientometrics. Main path analysis and Multi-RPYS of the citation network are used to show the changes and continuities in Schubert's intellectual career. Diachronic and static decomposition of a document set can lead to different results, while the analytically distinguishable lines of research may overlap and interact over time, and intermittent.

**Introduction**

In different compositions, the five of us have worked for the past two years on developing Referenced Publication Year Spectroscopy or—abbreviated—RPYS. RPYS is a bibliometric method which can be used to analyze the historical origins of research fields or researchers. This method analyzes the cited references (CR) and especially the referenced publication years of a publication set. The field CR in the Science Citation Index and the other databases at the Web of Science (WoS) contain a number of subfields separated by commas: the name of the first author, publication year, the abbreviated journal title, volume and page numbers, and increasingly also the doi (digital object identifier) of the cited document. In the online version (*SciSearch*) of the Science Citation Index at STN,[1] one can use these subfields for searching and retrieval (Marx, 2011; cf. Leydesdorff & Goldstone, 2014).

The first demonstration of RPYS as a method (Marx *et al*., 2014) was based on *SciSearch* at STN. In order to develop software for thus analyzing downloads from WoS, Lutz Bornmann linked up with Loet Leydesdorff, who extended his already existing software packages for bibliometric coupling[2] to this end (Leydesdorff *et al*., 2014). Andreas Thor further developed the program RPYS.exe (available at http://www.leydesdorff.net/software/rpys ) into the Cited References Explorer (at http://crexplorer.net; Thor *et al*., in preparation). CRExplorer not only allows for RPYS, but also includes a tool for the disambiguation of misspelled references. Comins & Hussey (2015a and b; Comins & Leydesdorff, in preparation) further developed RPYS into a tool for Multi-RPYS (available at http://comins.leydesdorff.net ). The occasion of a

---

[1] STN (or Science and Technology Networks) is a fee-based host of databases maintained by the American Chemical Society.
[2] The program BibJourn.exe (available at http://www.leydesdorff.net/software/bibjourn) uses the subfield of the abbreviated journal name for mapping the knowledge bases of document sets (e.g., Leydesdorff & Goldstone, 2014).

*Festschrift* for András Schubert's 70[th] birthday provides us with an opportunity to discuss the different options for RPYS in relation to the longer-term research program of algorithmic historiography—formulated by Garfield *et al.* (1964)—using Schubert's publications and citations as a bibliographic domain.

Garfield and Pudovkin further developed HistCite™ for the graphical user interfaces provided on both Windows and Apple computers in the late 1990s (Garfield *et al*., 2003; cf. Leydesdorff, 2010). The new version of HistCite™ (available at http://interest.science.thomsonreuters.com/forms/HistCite/) allows also for exporting the citation network into the Pajek format for social network analysis.[3] Hummon & Doreian (1989; Carley *et al*., 1993) first developed "main path analysis" that was integrated into Pajek in the 1990s. We will also pay attention to CitNetExplorer made available (at http://www.citnetexplorer.nl/) by researchers of the Center for Science and Technology Studies CWTS in Leiden (van Eck & Waltman, 2014) for citation network analysis.

**Data**

Searching for "AU = Schubert A and CI= Budapest", one retrieves 176 documents within the WoS domain of the Science and Social Science Citation Indices. Four of these documents are false positives (of Alfred Schubert).[4] We use the remaining 172 publications as our domain, downloaded on January 4, 2016.[5]

---

[3] Pajek is a program for network analysis and visualization, freely available for non-commercial purposes at http://mrvar.fdv.uni-lj.si/pajek/ .

---

[4] Four more papers can be added if conference proceedings are also taken into account; seven more documents were published during his doctorate period at the Physics Department of the University for Agricultural Sciences in Gödöllo. We are grateful to Wolfgang Glänzel for noting these corrections.
[5] Among these papers 20 are bibliographies and two meeting abstracts.

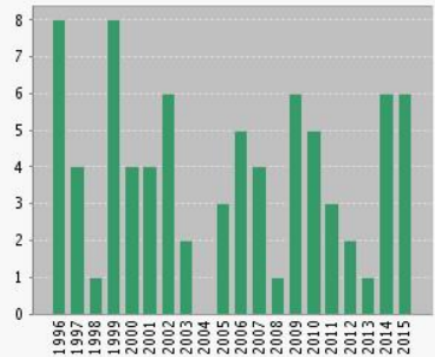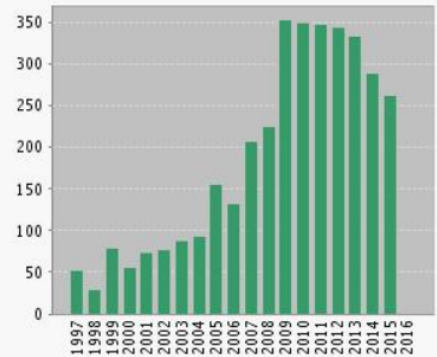

**Figure 1**: The Web of Science Citation Report for the 172 journal articles of András Schubert (January 18, 2016).

The WoS Citation Report in Figure 1 shows the publication and citation pattern of this set during the last twenty years. The legends show, among other things, that Schubert's papers are *on average* cited more than 24 times.

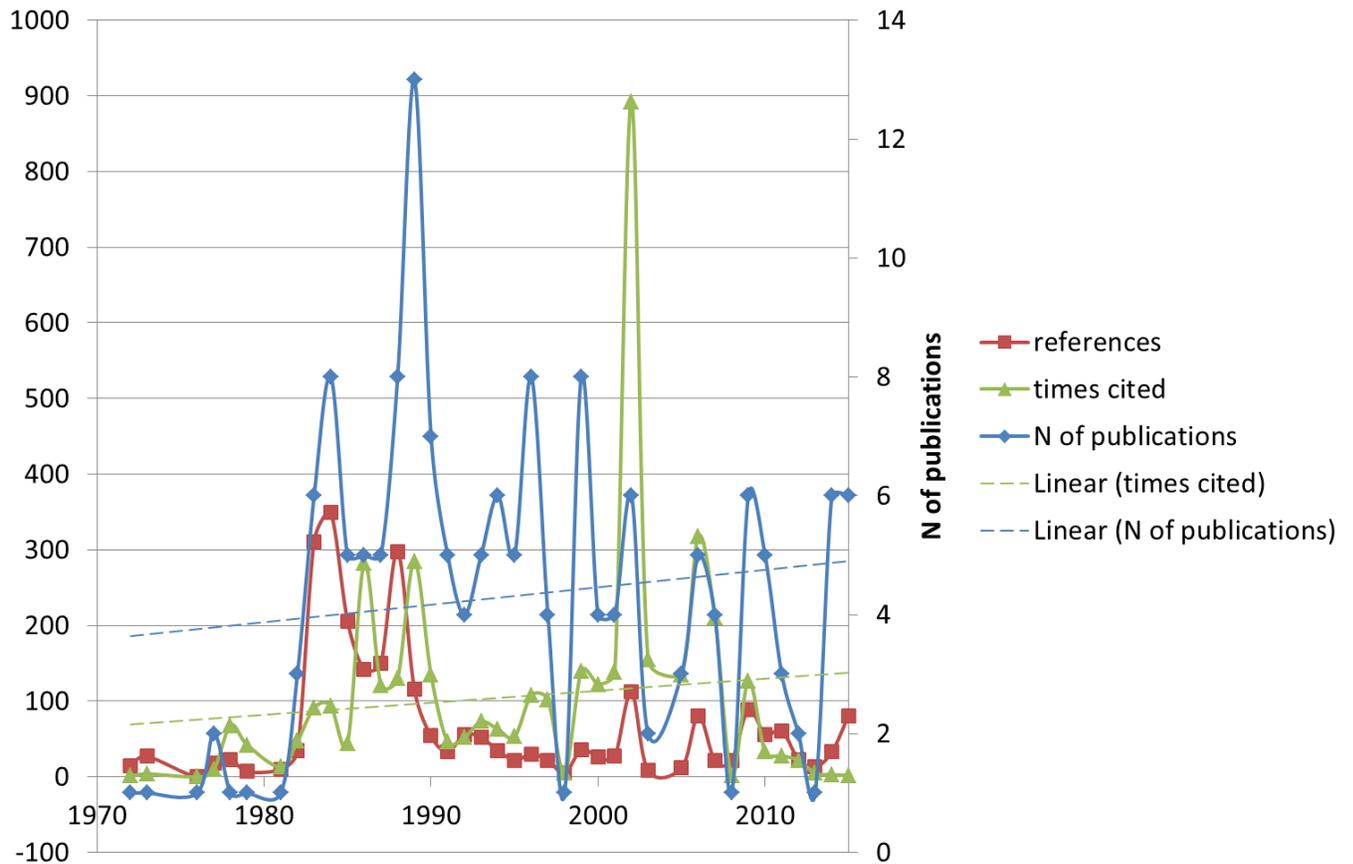

**Figure 2:** Publication, citation, and cited reference profiles of András Schubert, 1972-2015.

Figure 2 extends the graphs for the entire period 1972-2015. It shows the annual numbers of publications, citations, and cited references. As can be expected for a single author, publication and citation patterns fluctuate strongly over the entire period (if only for reasons of chance). Yet, both trends are upward as the dotted (regression) lines in Figure 2 reveal; there is a peak for publications in 1989 (n = 13) and for citations in 2002 (n = 892). Referencing is highest during the second half of the 1980s—the years of the establishment of the Information Science and Scientometrics Research Unit (ISSRU) in Budapest in collaboration with Tibor Braun and Wolfgang Glänzel. The total number of references by the 172 publications is 2,715; that is, 15.8 references per publication on average. Citation peaks in 2002 with 892 citations in WoS during that year. This peak is largely due to Schubert's coauthorship of a single publication (Barabási *et al.*, 2002) that has been cited 784 times.

**Algorithmic historiography**

*a. HistCite*

As mentioned above, Eugene Garfield's original program for algorithmic historiography was revived and further elaborated by Alexander Pudovkin when graphical interfaces became available on Windows computers in the late 1990s (Garfield *et al.*, 2003). HistCite™ is nowadays available upon registration at

http://interest.science.thomsonreuters.com/forms/HistCite/.[6]

---

[6] Using HistCite, the header of an input file—downloaded from WoS—needs to be changed from "FN Thomson Reuters Web of Science™" (the current header) into "FN ISI Export Format" (the old format) before HistCite can read the file. Under Microsoft Windows, HistCite requires the presence of the Internet Explorer. The input has to be saved as ASCII/ANSI.

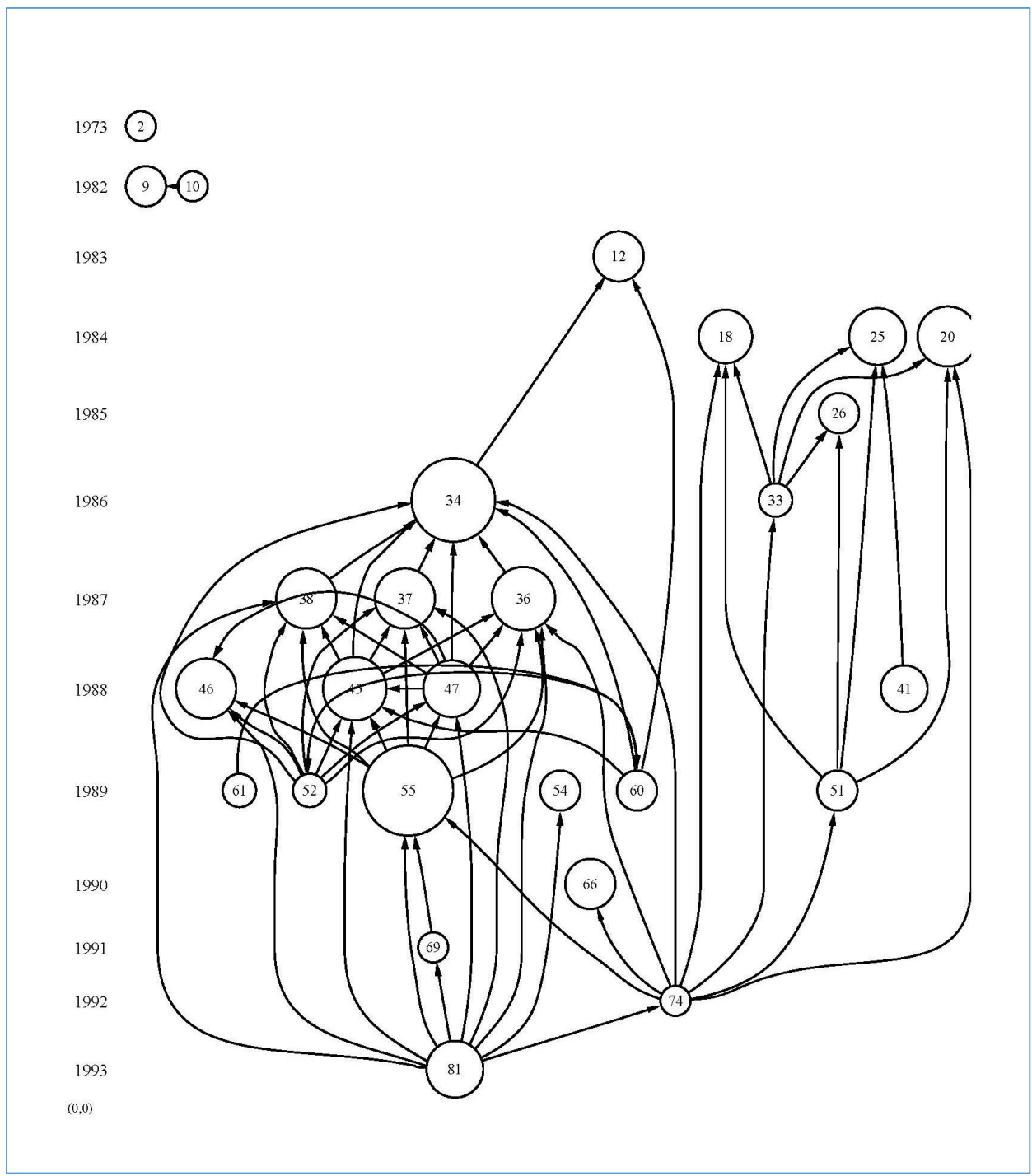

**Figure 3**: Default output of HistCite on the basis of 172 documents authored by András Schubert. The figure shows the top layer (n = 30) in the internal ("local") citation structure of his œuvre.

Figure 3 shows the HistCite network based on the so-called "Local Citation Scores" *within* the publication set of András Schubert. An alternative representation can be obtained by using the Global Citation Scores which are based on the times-cited scores in the input file. Since all input records are (co-)authored by Schubert, this figure shows the top-30 layer (n = 30) in his œuvre.[7] Self-citations to papers from the period 1983-1993 are prevalent in the set.

**Table 1**: Thirty papers selected by HistCite as the local citation network within Schubert's œuvre.
(LCS: Local Citation Score within this network; GCS: Global Citation Score using times-cited values).

| Nr in Fig. 2 | *Cited Reference* | LCS | GCS |
|---|---|---|---|
| 2 | NOSZTICZ.Z, 1973, PERIOD POLYTECH CHEM, V17, P165 | 2 | 4 |
| 9 | ZSINDELY S, 1982, SCIENTOMETRICS, V4, P57 | 4 | 26 |
| 10 | ZSINDELY S, 1982, SCIENTOMETRICS, V4, P69 | 2 | 21 |
| 12 | SCHUBERT A, 1983, SCIENTOMETRICS, V5, P59 | 6 | 62 |
| 18 | SCHUBERT A, 1984, J RADIOANAL NUCL CH, V82, P215 | 7 | 9 |
| 20 | SCHUBERT A, 1984, SCIENTOMETRICS, V6, P149 | 9 | 33 |
| 25 | GLANZEL W, 1984, Z WAHRSCHEINLICHKEIT, V66, P173 | 8 | 33 |
| 26 | TELCS A, 1985, MATH SOC SCI, V10, P169 | 4 | 10 |
| 31 | SCHUBERT A, 1986, CZECH J PHYS, V36, P121 | 2 | 27 |
| 32 | SCHUBERT A, 1986, CZECH J PHYS, V36, P126 | 4 | 21 |
| 33 | SCHUBERT A, 1986, SCIENTOMETRICS, V9, P231 | 3 | 18 |
| 34 | SCHUBERT A, 1986, SCIENTOMETRICS, V9, P281 | 16 | 215 |
| 36 | BRAUN T, 1987, SCIENTOMETRICS, V11, P9 | 10 | 30 |
| 37 | BRAUN T, 1987, SCIENTOMETRICS, V11, P127 | 9 | 24 |
| 38 | BRAUN T, 1987, SCIENTOMETRICS, V12, P3 | 9 | 22 |
| 41 | GLANZEL W, 1988, J INFORM SCI, V14, P123 | 5 | 37 |
| 45 | BRAUN T, 1988, SCIENTOMETRICS, V13, P181 | 10 | 43 |
| 46 | BRAUN T, 1988, SCIENTOMETRICS, V14, P3 | 9 | 28 |
| 47 | BRAUN T, 1988, SCIENTOMETRICS, V14, P365 | 8 | 18 |
| 51 | SCHUBERT A, 1989, J AM SOC INFORM SCI, V40, P291 | 4 | 12 |
| 52 | BRAUN T, 1989, SCIENTOMETRICS, V15, P13 | 3 | 6 |
| 54 | BRAUN T, 1989, SCIENTOMETRICS, V15, P325 | 4 | 21 |
| 55 | SCHUBERT A, 1989, SCIENTOMETRICS, V16, P3 | 19 | 186 |
| 60 | BRAUN T, 1989, TRAC-TREND ANAL CHEM, V8, P281 | 4 | 14 |

---

[7] See for further explanation of the definitions in HistCite at http://garfield.library.upenn.edu/histcomp/guide.html.

| | | | |
|---|---|---|---|
| 61 | BRAUN T, 1989, TRAC-TREND ANAL CHEM, V8, P316 | 3 | 7 |
| 66 | SCHUBERT A, 1990, SCIENTOMETRICS, V19, P3 | 6 | 108 |
| 69 | BRAUN T, 1991, SCIENTOMETRICS, V20, P9 | 2 | 6 |
| 70 | SCHUBERT A, 1991, SCIENTOMETRICS, V20, P317 | 2 | 12 |
| 74 | SCHUBERT A, 1992, SCIENTOMETRICS, V23, P3 | 2 | 13 |
| 81 | BRAUN T, 1993, SCIENTOMETRICS, V28, P137 | 8 | 20 |

HistCite provides a legend in a separate (split) screen (Table 1). Node 34, for example, is self-cited eight times; this paper, coauthored with Tibor Braun (Schubert & Braun, 1986), seems to have been constitutive for the research program thereafter.

HistCite can also be used to generate a complete citation network by setting the limit above the size of the set under study (instead of the 30 which are the default for the graph in HistCite, for example, 172 in our case). This network is exported in the Pajek (.net) format that can be used in many network analysis and visualization programs such as UCInet, Gephi, and VOSviewer. Pajek furthermore offers the option to study the main path in the network.

*b. Analysis and visualization of the citation network*

The network file exported from HistCite contains the 172 documents as nodes and the citation relations among them as links; 95 nodes are linked into a largest component. This largest component can be visualized as a citation network (Figure 4). By choosing the layout of Fruchterman & Reingold (1992), we can observe the two constitutive clusters of the ISSRU research program to the left in the bottom half. One cluster is dominated by papers with Tibor Braun as first author and the other by papers with András Schubert as first author. Wolfgang Glänzel joined the Budapest group first as a PhD student and then became a third (co-)author in the second half of the 1980s. Most of the papers are coauthored by at least two of these three authors.

At the top right of Figure 4, one can see that the recent work of Schubert (since 2005) is only weakly related to earlier work in terms of citation relations; references to papers coauthored with Glänzel as lead author provide the relationship with evaluation studies. In 2005, Jorge Hirsch published his study of the *h*-index which opened a whole new set of questions for bibliometric investigation. Thirteen of the 44 papers in the period 2005-2015 (that is, 30%) contain the words "h-index" or "Hirsch" in the *title*. Within this cluster of most recent papers, Tibor Braun is the lead author in two cases.

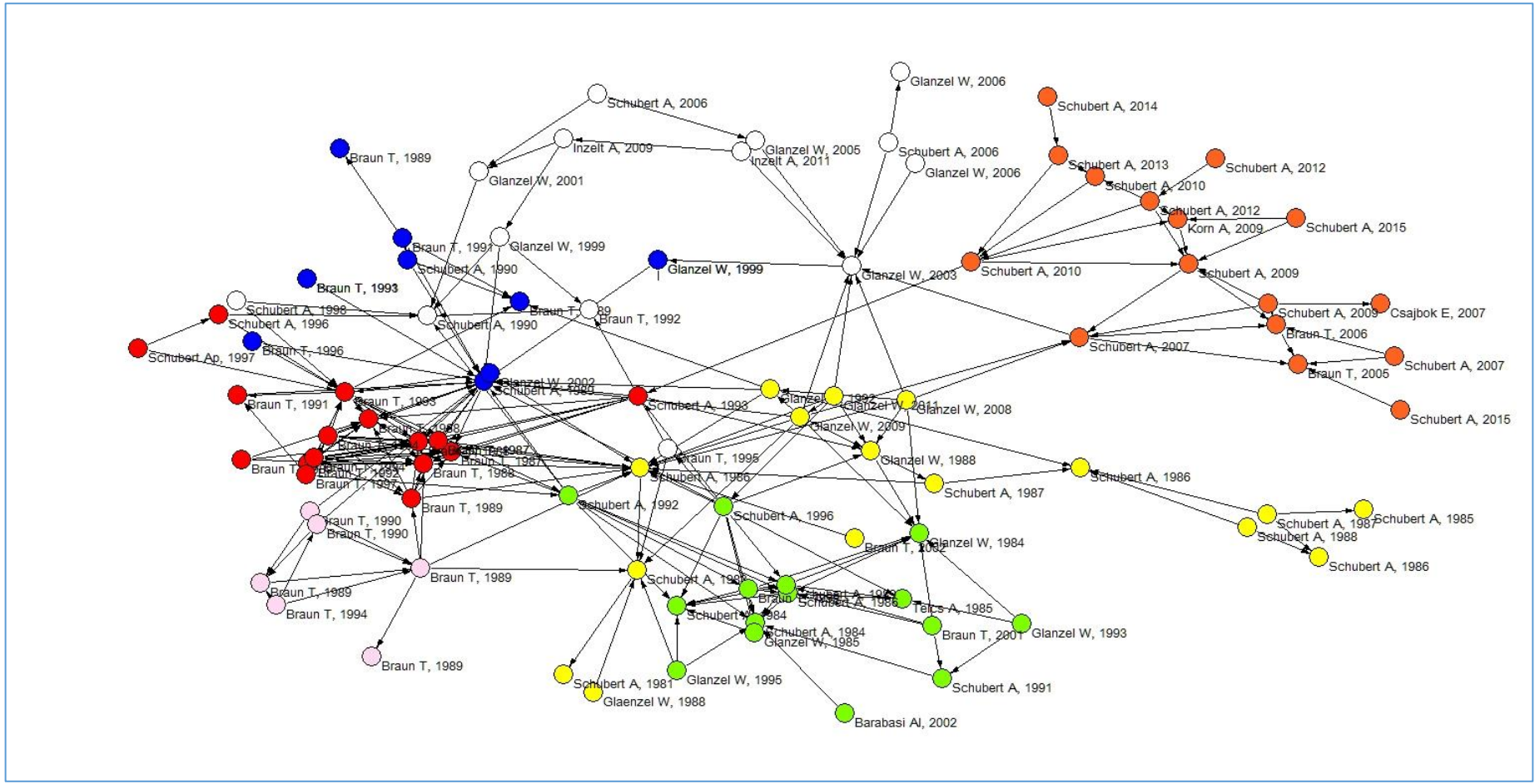

**Figure 4**: Seven clusters in the main component of the citation matrix (n = 95), distinguished using Blondel *et al*. (2008) in Pajek; Fruchterman & Rheingold (1992) was used for the layout.

One advantage of network analysis and visualization programs is the availability of algorithms for the decomposition and further statistics, whereas HistCite™ has remained descriptive. In Figure 4, for example, seven clusters were distinguished by using the decomposition algorithm of Blondel *et al*. (2008). The modularity $Q$—a measure for the dividedness between 0 and 1—of the network is 0.578. Thus, the clusters are weakly distinct. Similarly, one can feed the Pajek file into VOSviewer and obtain a comparable network. The algorithm then reveals a finer distinction of 11 clusters in the large component, and one obtains other options for the visualization (not shown here). More specifically developed for citation network analysis is the program CitNetExplorer of the same group at the Center for Science and Technology Studies CWTS (van Eck & Waltman, 2014).

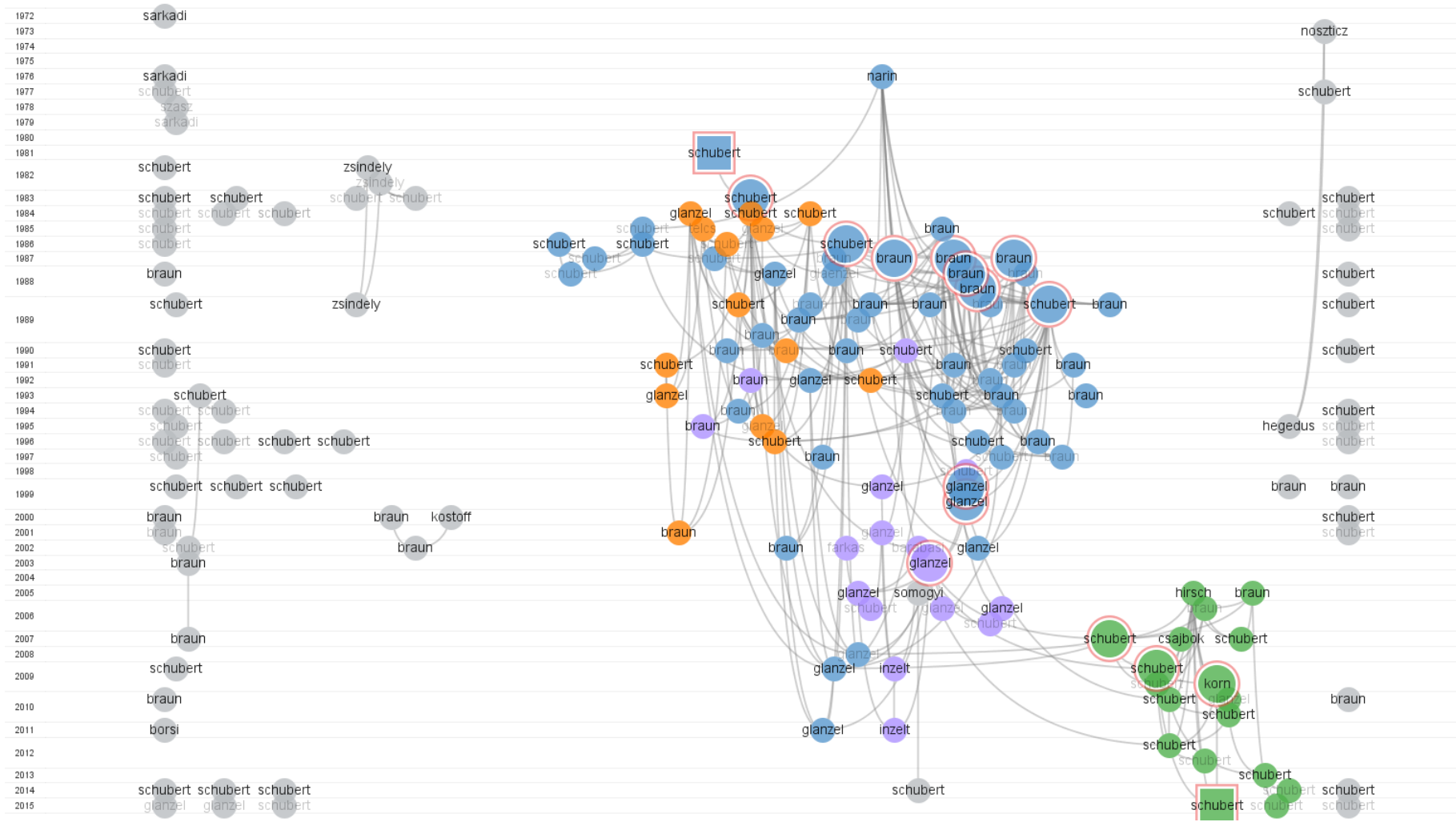

**Figure 5**: Clustering and indication of shortest path applying CitNetExplorer to the citation network of Schubert's œuvre (at

http://www.citnetexplorer.nl/ ).

Figure 5 shows the results of feeding the original WoS download (n = 172) into CitNetExplorer. By default, the program analyzes only the articles with a times-cited score equal to or larger than ten. As against the default of making only 40 nodes (papers) visible, we chose to make all the remaining papers visible. This includes a number of papers which are not connected and therefore colored grey in Figure 5.

The clustering algorithm of CitNetExplorer distinguishes four main groups with a minimum size of ten. One of them is the recent group of papers (colored green) and is discrete from the other three which are more mixed. Although one can distinguish the Braun-dominated cluster from the Schubert-dominated one during the late 1980s and 1990s, the division is fuzzy. The third group in the first decade of the 2000s is dominated by Glänzel's papers (lilac). The visualization of CitNetExplorer not only labels with the citing papers, but includes the cited first authors; for example, Hirsch (2005).[8]

Within CitNetExplorer, the analyst can mark two nodes and ask for the shortest path between them. In Figure 5, we marked Schubert & Braun (1981) as the first paper in the common cluster, and Schubert (2015) at the bottom as the last paper. These two nodes are marked on the map with (orange) squares. More than a single shortest path (in six steps) was reported in this data. The papers on a shortest path are indicated with orange circles around the nodes.

---

[8] Insofar as the cited references are to citing papers in the set, the title-field is imported into the documentation of the visualization.

*b. Main path analysis in Pajek*

Unlike the shortest path between two nodes selected by the analyst, the main path is defined as a systemic property. Citing previous literature and being cited by subsequent literature position a paper in relation to other papers in the set (Hummon & Doreian, 1989). When a set of documents represents a self-contained field—not significantly building on knowledge from other fields—the citation network among the key documents (the most highly cited ones) can be expected to contain at least one main path (Carley *et al*., 1993). Main-path algorithms enable us to make the structural backbone of a literature visible (Lucio-Arias & Leydesdorff, 2008).

The main path is reconstructed by calculating the connectivity of the links in terms of their degree centrality and outlining the path formed by the nodes with the highest degree. In terms of a citation network, this degree measure considers the number of citations a document receives (indegree) as well as the number of cited references in the documents (outdegree). The main path is constructed by selecting those connected documents with the highest scores until an end document is reached (Batagelj, 2003). This can be either a document that is no longer cited or one that contains no further references within the set.

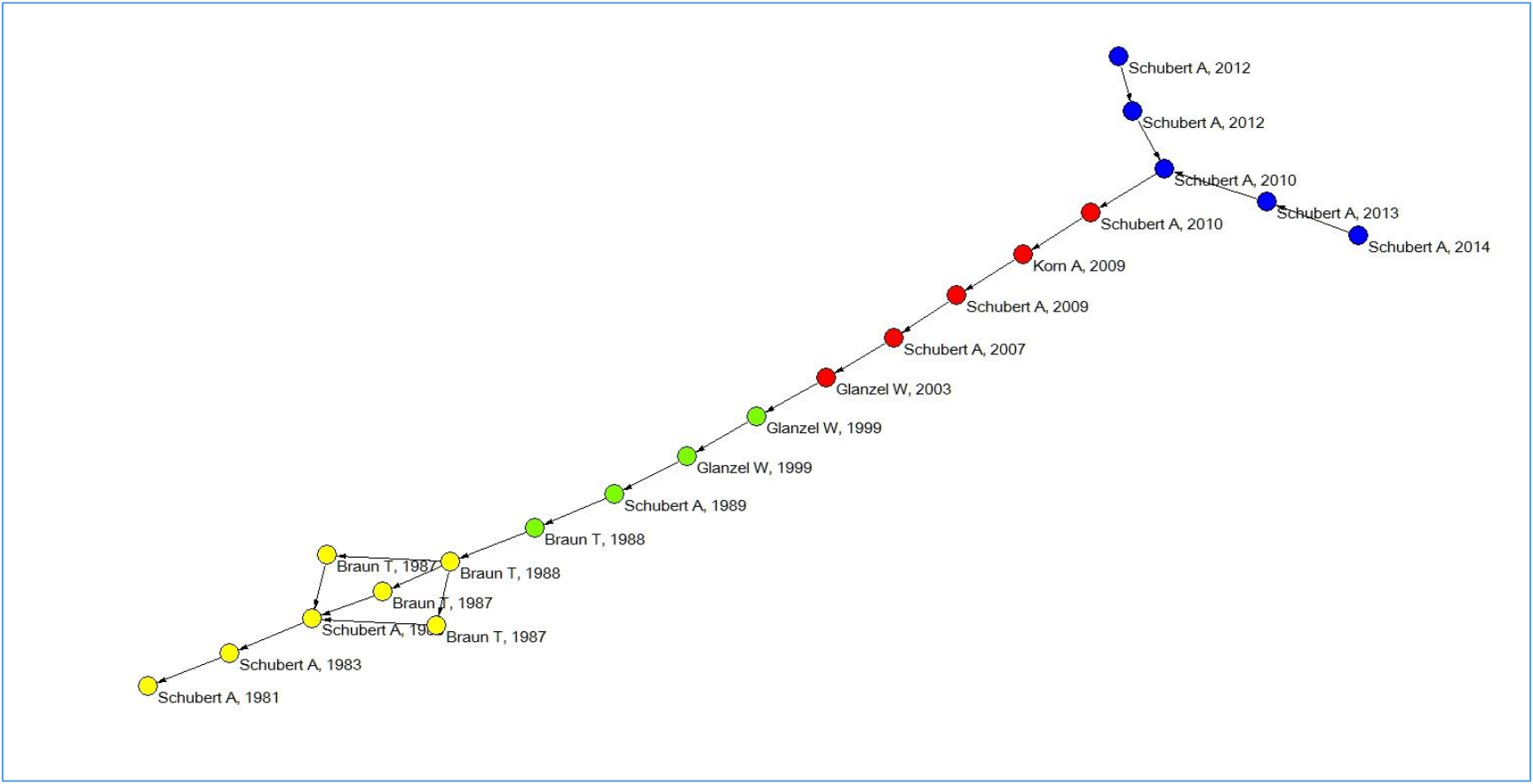

**Figure 6**: Main path in the citation network of András Schubert's œuvre. Blondel *et al.* (1998) was used for the decomposition and Kamada & Kawai (1989) for the layout.

The main path shown in Figure 6 can be extracted in Pajek as a partition from the citation network. Although we did not add the years as a temporal dimension to the documents (as in the shortest path analyzed above), the algorithm itself sorts the references along a time line. Using Blondel's *et al*. (1998) algorithm for the decomposition, four clusters are robustly indicated ($Q$ = 0.588).[9] The first cluster (yellow) shows the initial period of institutionalization of the ISSRU unit and the journal *Scientometrics* during the 1980s. The second period represents the 1990s; the third (red) the period begins after Glänzel left the unit for Louvain in 2002. Schubert himself, however, begins new research lines since 2010. These latter papers are all first-authored by him, whereas in the previous periods coauthorship with Glänzel was also common on the main path.

Note that these are distinctions within the construct of the main path. They inform us about the network structure of citation relations, potentially including relations among different research lines. We refrain from rationalizing the transitions indicated in Figure 6 in terms of intellectual changes, but return to this issue more extensively in the discussion section.

---

[9] We formulate "robustly" because this analysis can be repeated.

*d. RPYS*

RPYS plots the cumulative distribution of cited references in terms of the referenced publication years. The peaks in the graph are often discrete and thus indicate specific publications which were highly cited within the domain of the sample. But this is not the case at the research front—that is, the most recent years—because the citation classics are not yet sorted out in that part of the domain (Price, 1970). Baumgartner & Leydesdorff (2014) distinguish between transient knowledge claims at the research front and sticky ones which remain highly cited after more than ten years. One can also consider the latter citations as concept-symbols (Small, 1978) and the former as citation currency.

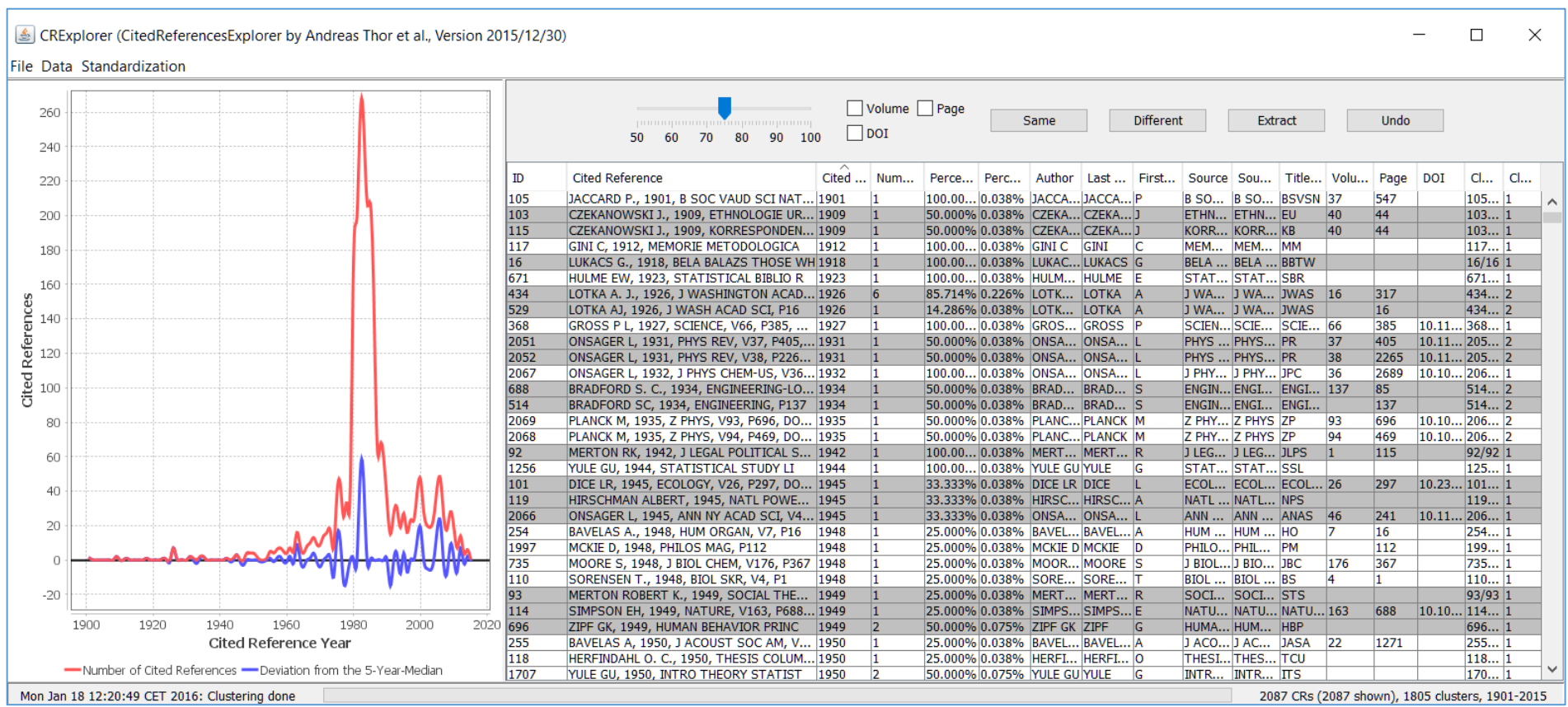

**Figure 7**: User interface of CRExplorer after importing the œuvre of Schubert.

Figure 8 shows the results of using CRExplorer for RPYS. The red line visualizes the number of cited references per referenced publication year during the period 1900-2015. In order to identify those publication years with significantly more cited references than other years, the deviation of the number of cited references in each year from the median of the number of cited references in the two previous, the current, and the two following years ($t – 2; t – 1; t ; t + 1; t + 2$) is visualized as a blue line. This deviation from the five-year median provides a smoother curve than one in terms of absolute numbers.

The disadvantage of the figure in the left pane is the possibility that several papers may be adding up to a peak in a specific year. Inspection of the listing in the right pane teaches us that the first peak in the figure to the publication year 1926 points to Lotka (1926), which is a citation classic in this field; but the 1963 peak, for example, is composed of several classics: Price (1963), Irwin (1963), and Kessler (1963), cited four, three, and three times, respectively. Furthermore, Lotka (1926) is cited six times as "LOTKA A. J., 1926, J WASHINGTON ACAD SC, V16, P317", but also once as "LOTKA AJ, 1926, J WASH ACAD SCI, P16".

Although Thomson Reuters standardizes the cited references of papers included in WoS, the problem of variants of the same cited references remains, potentially disturbing the results of RPYS and citation analysis more generally. If cited references are available with several variants, it is no longer possible to produce a reliable list or ranking of the most frequently cited publications. Evaluation studies are very susceptible to this type of error. The problem of variants is especially urgent for document types other than journal papers (such as books and book chapters). Can the cited references be disambiguated?

CRExplorer offers the possibility to cluster the variants of cited references. A detailed description of the clustering and merging methods used in the program can be found in Thor *et al*. (2016, in preparation). After a first round of automatic cleaning, one can proceed with manual cleaning. Since the automatic clustering of variants can also be a source of error, one is advised to check and possibly correct the results of the automatic clustering manually. Note that not all errors can be corrected because references may be incomplete (Leydesdorff, 2008: 285, Table 4).

**Table 2**: After disambiguation (CRExplorer), Glänzel (1988) is added to the publications referenced more than five times in the set; Narin (1976) and Braun (1987) are ranked at a higher position.

| CR | LCS |
| --- | --- |
| BRAUN T, 1985, SCIENTOMETRIC INDICA | 24 |
| SCHUBERT A, 1989, SCIENTOMETRICS, V16, P3 | 19 |
| SCHUBERT A, 1986, SCIENTOMETRICS, V9, P281 | 16 |
| HIRSCH JE, 2005, P NATL ACAD SCI USA, V102, P16569 | 13 |
| BRAUN T, 1987, SCIENTOMETRICS, V11, P9 | 10 |
| BRAUN T, 1988, SCIENTOMETRICS, V13, P181 | 10 |
| **NARIN F., 1976, EVALUATIVE BIBLIOMET** | **10** |
| **BRAUN T, 1987, LIT ANAL CHEM SCIENT** | **9** |
| BRAUN T, 1987, SCIENTOMETRICS, V11, P127 | 9 |
| BRAUN T, 1987, SCIENTOMETRICS, V12, P3 | 9 |
| BRAUN T, 1988, SCIENTOMETRICS, V14, P3 | 9 |
| SCHUBERT A, 1984, SCIENTOMETRICS, V6, P149 | 9 |
| BRAUN T, 1988, SCIENTOMETRICS, V14, P365 | 8 |
| BRAUN T, 1993, SCIENTOMETRICS, V28, P137 | 8 |
| GLANZEL W, 1984, Z WAHRSCHEINLICHKEIT, V66, P173 | 8 |
| GLANZEL W, 2003, SCIENTOMETRICS, V56, P357 | 8 |
| GARFIELD E, 1972, SCIENCE, V178, P471 | 7 |
| PRICE DJD, 1976, J AM SOC INFORM SCI, V27, P292 | 7 |
| SCHUBERT A, 1984, J RADIOANAL NUCL CH, V82, P215 | 7 |
| **GLANZEL W, 1988, J INFORM SCI, V14, P123** | **6** |
| HAITUN SD, 1982, SCIENTOMETRICS, V4 | 6 |
| HAITUN SD, 1982, SCIENTOMETRICS, V4, P5 | 6 |
| HAITUN SD, 1982, SCIENTOMETRICS, V4, P89 | 6 |
| IRWIN JO, 1975, J ROY STAT SOC A STA, V138, P18 | 6 |
| IRWIN JO, 1975, J ROY STAT SOC A STA, V138, P204 | 6 |
| IRWIN JO, 1975, J ROY STAT SOC A STA, V138, P374 | 6 |
| LOTKA A. J., 1926, J WASHINGTON ACAD SC, V16, P317 | 6 |
| SCHUBERT A, 1983, SCIENTOMETRICS, V5, P59 | 6 |
| SCHUBERT A, 1990, SCIENTOMETRICS, V19, P3 | 6 |
| TAGUE J, 1981, J AM SOC INFORM SCI, V32, P280 | 6 |

Table 2 lists the publications referenced more than five times by András Schubert's publication set after careful (automatic and manual) clustering of the cited references using CRExplorer. Two publications change positions in the hierarchy, and one (Glänzel & Schubert, 1988) was added to the set of 29 publications referenced more than five times. Francis Narin's (1976) book on the use of bibliometrics in evaluation, for example, is referenced with four variants. It is cited ten instead of seven times in the publications of András Schubert after the disambiguation process.

*f. Multi-RPYS*

Multi-RPYS provides an extension of standard RPYS methodology and was developed to make possible comparative analysis among different years and/or different sets. This objective is accomplished by applying a rank-transformation to the standard RPYS outputs and by visualizing the results as a heat map. Multi-RPYS has hitherto been used to investigate (1) communal intellectual histories across different journals, and (2) the temporal dynamics of historical influences (Comins & Hussey, 2015a; Comins & Hussey, 2015b; Comins & Leydesdorff, in preparation). Specifically, this latter technique segments the set of citing articles by publication year and generates a Multi-RPYS heat map across these segments to track *when* and *how consistently* references were cited by researchers. Below we use this approach to consider shifts in the intellectual influences driving András Schubert's œuvre.

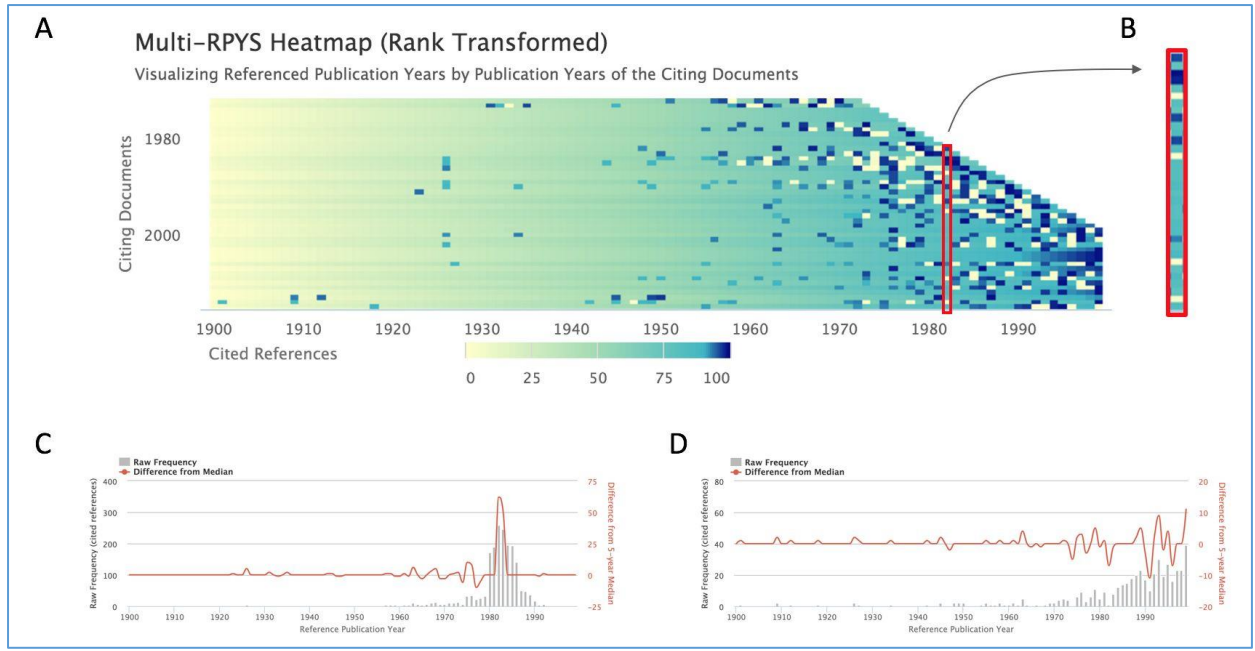

**Figure 8**. Multi-RPYS heatmap computing RPYS results for Schubert's œuvre segmented by publications year of the citing documents (along the y-axis).

The largest peak in the RPYS plot of Schubert's works occurs in 1982 (see Figure 7), and is driven by Haitun's (1982) three papers about "Stationary Scientometric Distributions" published as different parts in *Scientometrics*. The band (B) in Figure 9 corresponds to 1982 as the referenced publication year. It shows that most citations to this year occurred from citing documents—chronologically sorted along the y-axis—published in the first half of Schubert's career. By splitting (in the lower part of the figure) the works of Schubert into those published from 1972-1993 (C) and 1994-2015 (D), the absence of 1982 as a peak reference year in the latter set becomes visible. In other words, Haitun's work was cited by Schubert only during the first part of his career.

**RPYS and bibliographic coupling**

The data used for RPYS and citation network studies (CR) can also be used for bibliographic coupling (Kessler, 1963). What is the difference? In citation network studies and RPYS, cited references across the sets under study are binned in years; in studies of bibliographic coupling one uses the citing documents as units of analysis. Using years, heterogeneous sets in terms of cognitive contents and social relations are potentially lumped together. Figure 10, for example, shows the clear structure that can instead be found in Schubert's œuvre when these same cited references are used for a map of the bibliographic coupling among the co-authors of Schubert.

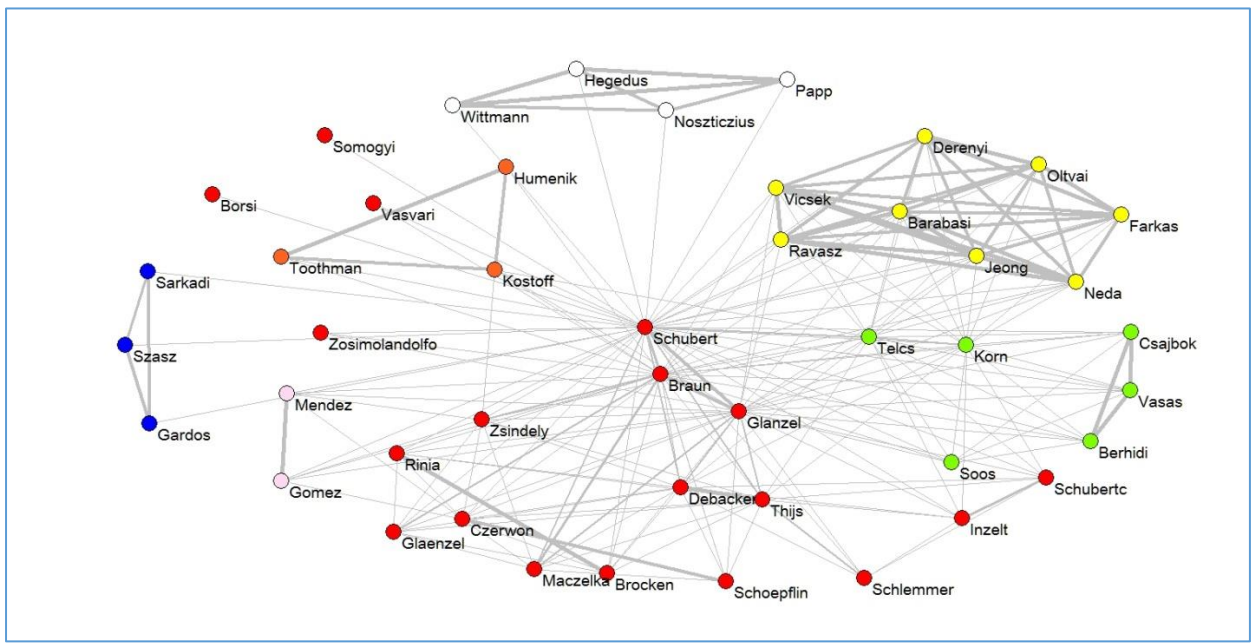

**Figure 10**. Bibliographic coupling of 44 co-authors of Schubert's 172 publications; seven clusters were distinguished by the algorithm of Blondel *et al*. (2008); $Q = 0.639$; Kamada & Kawai (1989) was used for the visualization; the output is cosine-normalized.

We shall not discuss Figure 10 here; but show it in order to make the point that diachronic analysis and static analysis can lead to very different results. One cannot easily map the relations among 44+1 (co-)authors diachronically. Using a dynamic optimization among multi-dimensional scaling outputs for subsequent years, however, Leydesdorff & Schank (2008) have developed a version of visone[10] (visone v2.3.X at http://www.leydesdorff.net/visone) that allows for combining social and cognitive attributes of documents in animations (e.g., Leydesdorff, 2010a and b).

One disadvantage of focusing on cited references in terms of referenced publication years is the neglect of the knowledge content which structures citation networks in the development of the sciences. One risks studying the dynamics of citations instead of the dynamics of science. Combining the referenced publication years with the cited journals may provide a perspective for the further development of Multi-RPYS in a direction that will show the development of socio-cognitive structures in the data over time (cf. Leydesdorff & Goldstone, 2014).

**Conclusions**

RPYS is a recently introduced method for the study of the historical roots of research fields or researchers. It is based on the analysis of cited references and especially cited reference years. The occasion of a *Festschrift* for András Schubert's 70[th] birthday provides us with the opportunity to discuss the different options for RPYS in relation to the longer-term research program of algorithmic historiography using Schubert's publications and the references cited therein as a bibliographic domain. The results show that RPYS allows for the reconstruction of the shoulders on which a researcher stands. Without disambiguation, however, the CR field remains an unreliable source. Using it for evaluation purposes requires disambiguation. CRExplorer offers a partial solution to this problem.

---

[10] Visone is a network analysis and visualization program, freely available at http://visone.info .

The largest peak in the RPYS plot of Schubert's publications (which indicates the works with the largest influence on Schubert's research) occurs in 1982, and is driven by Haitun's (1982) three papers about "Stationary Scientometric Distributions". The results of Multi-RPYS revealed, however, that Haitun's papers were primarily referenced during the first half of Schubert's career. These and further results in this study based on András Schubert's publications demonstrate that RPYS is a useful addition to the already available bibliometric techniques for algorithmic historiography (such as *HistCite*™, CitNetExplorer, visone, etc.).

**Acknowledgement**

We are grateful to Wolfgang Glänzel for his comments on an earlier version of this paper.